\documentclass[prd,twocolumn,floatfix,nofootinbib,superscriptaddress,
showpacs]{revtex4}
\usepackage[utf8]{inputenc}
\usepackage{graphicx}
\usepackage{epsfig}
\usepackage{bm}
\usepackage{amssymb}
\usepackage{float}
\usepackage{amsmath}
\usepackage{dcolumn}
\usepackage{tensor}
\usepackage{cancel}
\usepackage[colorlinks]{hyperref}
\usepackage[usenames,dvipsnames]{color}
\hypersetup{
     breaklinks=true,
    pdfstartview={FitH},    
    colorlinks=true,       
    linkcolor=blue,          
    citecolor=red,        
    filecolor=magenta,      
    urlcolor=blue,           
    anchorcolor=green,      
    linktocpage=true
}

\def\a{ \alpha}
\def\b{ \beta}

\def\d{ \delta}
\def\m{ \mu}
\def\n{ \nu}

\def\r{ \rho}

\def\a{ \alpha}
\def\b{ \beta}

\def\d{ \delta}
\def\m{ \mu}
\def\n{ \nu}

\def\r{ \rho}
\def\k{ \kappa}

\def\pm{p_m}
\def\rm{\rho_m}
\def\hd{\dot{H}}

\def\be{\begin{equation*}}
\def\ee{\end{equation*}}

\usepackage{amsmath,latexsym}
\usepackage{placeins}
\usepackage[toc,page]{appendix}

\newcommand{\mathsym}[1]{{}}
\newcommand{\unicode}[1]{{}}

\newcommand{\bea}{\begin{eqnarray}}
\newcommand{\eea}{\end{eqnarray}}
\newcommand{\beaa}{\begin{eqnarray*}}
\newcommand{\eeaa}{\end{eqnarray*}}

\def\arxiv{http://arxiv.org/abs}

\begin{document}
\title{Modified gravity and cosmology with nonminimal (derivative) coupling 
between  matter and the Einstein tensor }

\author{Petros Asimakis}
\affiliation{Department of Physics, National Technical University of Athens, 
Zografou Campus GR 157 73, Athens, Greece}

\author{Spyros Basilakos}
\affiliation{National Observatory of Athens, Lofos Nymfon, 11852 Athens, 
Greece}
\affiliation{Academy of Athens, Research Center for Astronomy and Applied 
Mathematics, Soranou Efesiou 4, 11527, Athens, Greece}
\affiliation{School of Sciences, European University Cyprus, Diogenes 
Street, Engomi, 1516 Nicosia, Cyprus}

\author{Andreas Lymperis}
\affiliation{Department of Physics, University of Patras, 26500 Patras, Greece}

 \author{Maria Petronikolou}
\affiliation{Department of Physics, National Technical University of Athens, 
Zografou Campus GR 157 73, Athens, Greece}

\author{Emmanuel N. Saridakis}
\affiliation{National Observatory of Athens, Lofos Nymfon, 11852 Athens, 
Greece}
\affiliation{CAS Key Laboratory for Researches in Galaxies and Cosmology, 
Department of Astronomy, University of Science and Technology of China, Hefei, 
Anhui 230026, P.R. China}
 \affiliation{Departamento de Matem\'{a}ticas, Universidad Cat\'{o}lica del 
Norte,  Avda. Angamos 0610, Casilla 1280 Antofagasta, Chile}

\begin{abstract}
We  construct new classes of modified theories in which the matter sector 
couples with the Einstein tensor, namely we consider direct couplings of the 
latter to the energy-momentum tensor, and to the derivatives of its  trace.
We extract  the general field equations, which do not contain higher-order 
derivatives, and we apply them in a cosmological 
framework, obtaining the Friedmann equations, whose extra terms give rise to 
an effective dark energy sector. At the background level we show  that we can 
successfully describe  the usual thermal history of the universe, with  the 
sequence of matter and dark-energy epochs, while  the dark-energy 
equation-of-state parameter can lie in the phantom regime, tending progressively 
to $-1$ at   present and future times.  
Furthermore, we confront  the theory with Cosmic Chronometer data, showing 
that  the agreement is very good. Finally, we perform
a detailed investigation of scalar and tensor perturbations, and extracting an 
approximate evolution equation for the matter overdensity we show that the 
predicted behavior is in agreement with observations.
\end{abstract}

\maketitle

\section{Introduction}

According to the concordance  model of cosmology the universe is currently  
accelerating, while it entered this era after   being in a long 
matter-dominated 
epoch. This behavior, alongside the early accelerated era of inflation, 
cannot be reproduced within the standard framework of general relativity and 
Standard Model of particles, and thus extra degrees of freedom should be 
introduced. One can attribute these extra degrees of freedom to new, exotic 
forms of matter, such as the inflaton field at early times (for reviews 
see \cite{Olive:1989nu,Bartolo:2004if}) and/or the dark energy concept at late  
times (for reviews see \cite{Copeland:2006wr,Cai:2009zp}). Alternatively, one 
can consider the extra degrees of freedom to have gravitational origin, i.e. to 
arise from a gravitational modification that possesses general relativity as a 
particular limit (see \cite{Nojiri:2006ri,Capozziello:2011et,Cai:2015emx} and 
references therein).

In order to construct gravitational modifications one usually starts from the 
Einstein-Hilbert Lagrangian and extends it in various ways, resulting in   
$f(R)$ 
gravity \cite{DeFelice:2010aj}, in  Gauss-Bonnet and $f(G)$ gravity
\cite{Antoniadis:1993jc,Nojiri:2005jg},  in   Lovelock and  f$($Lovelock$)$  
gravity \cite{Lovelock:1971yv,Deruelle:1989fj}, 
etc (for a review see \cite{CANTATA:2021ktz}). 
On the other hand, one may start  from the equivalent, torsional
formulation of gravity, and extend it in similar ways, obtaining 
  $f(T)$ gravity \cite{Bengochea:2008gz,Chen:2010va},  $f(T,B)$ gravity 
\cite{Bahamonde:2015zma},  $f(T,T_G)$ gravity
\cite{Kofinas:2014owa} etc. Nevertheless, one can consider theories in which 
the geometric part of the action is coupled to the non-geometric sector. In the 
case of general relativity the simplest models are those 
with non-minimally coupled 
\cite{Uzan:1999ch,deRitis:1999zn,Bertolami:1999dp,Faraoni:2000wk} and 
non-minimal-derivatively 
coupled 
\cite{Amendola:1993uh,Capozziello:1999xt,Daniel:2007kk,Saridakis:2010mf,
Sadjadi:2010bz,ArmendarizPicon:2000ah}
scalar fields, or the general scalar-tensor  
\cite{Jordan:1959eg,Brans:1961sx,Damour:1992we} and  
Horndeski/Galileon theories 
\cite{Horndeski:1974wa,DeFelice:2010nf,Deffayet:2011gz}. In the case of 
torsional gravity one can similarly construct   
scalar-torsion theories \cite{Geng:2011aj,Hohmann:2018dqh} or teleparallel 
Horndeski gravity  \cite{Bahamonde:2019shr,Bahamonde:2021dqn}.

Inspired by these couplings between geometric and non-geometric sectors, one 
can proceed to the construction of theories in which  the gravitational sector 
couples in a non-trivial way with the matter one, since  there is no 
theoretical 
reason against such interactions.
The simplest way is to consider that the matter Lagrangian $\mathcal{L}_m$ is
coupled to functions of the Ricci scalar 
\cite{Bertolami:2007gv,Bertolami:2008zh,Bertolami:2008ab}, which can be 
extended  to arbitrary functions of $\left(R, L_m\right)$  
\cite{Harko:2008qz,Harko:2010mv,Harko:2012hm,Wang:2012rw,Fadragas:2014mra}. 
Additionally,
one can consider models where the Ricci scalar is coupled to the trace of
the energy momentum tensor $T$ and extend to arbitrary functions,
such as in $f(R,T)$ theory 
\cite{Harko:2011kv,Momeni:2011am,Sharif:2012zzd,Alvarenga:2013syu,
Shabani:2013djy,Noureen:2015xva,Zaregonbadi:2016xna},
or even consider terms of the form  $R_{\mu\nu}T^{\mu\nu}$ 
\cite{Haghani:2013oma,Odintsov:2013iba}. Alternatively, one can follow the same 
path in the case of torsional gravity, and construct modifications in which the 
matter Lagrangian is coupled to  functions of the torsion scalar    
\cite{Harko:2014sja,Carloni:2015lsa,Gonzalez-Espinoza:2018gyl}, as well as 
theories where the torsion scalar is coupled to the trace of the energy momentum 
tensor   
\cite{Harko:2014aja,Junior:2015bva,Saez-Gomez:2016wxb,Farrugia:2016pjh,
Pace:2017dpu}. 
We mention here that  
the above modifications, in which one handles the gravitational and matter
sectors on equal footing, do not present any problem at the theoretical
level, and one would only obtain observational constraints only in the case of 
baryonic matter due to non-geodesic motion.

In the present work, inspired by the coupling of the scalar fields to the 
Einstein tensor, as well as by the coupling of the matter sector to the Ricci 
scalar, we are interested in constructing new classes of modified theories, in 
which the matter sector couples with the Einstein tensor. As one can see, we 
can directly couple the energy-momentum tensor to the Einstein tensor, i.e. 
consider a term $G_{\mu\nu} T^{\mu\nu}$, or we can couple the Einstein 
tensor to derivatives of the trace of the   energy-momentum tensor, i.e. 
consider a term $G_{\mu\nu}(\partial^{\mu}T)(\partial^{\nu}T)$. Interestingly 
enough, similarly to the other matter-gravity couplings, in the cosmological 
applications of these theories the extra terms in the Friedmann equations, 
although originating from the matter sector, can lead to accelerated expansion. 

The plan of the work is the following: In Section \ref{Theory} we present the 
theoretical basis of the theories with   couplings 
between matter sector and the Einstein tensor, extracting the general field 
equations. In Section \ref{Applications} we proceed to the cosmological 
application, and in particular in subsection \ref{Background} we investigate 
the background behavior, while in subsection \ref{perturbations} we perform a 
detailed 
perturbation analysis. Finally, in Section \ref{Conclusions}
we summarize the obtained results.

\section{Non-minimal  (derivative) couplings between matter and Einstein tensor}
\label{Theory}

In this work we propose more general couplings between the geometric and 
the matter sectors, and in particular we study actions in which the 
energy-momentum tensor and its trace couple to the Einstein tensor. We consider 
actions of the form 
\begin{eqnarray}
&& 
\!\!\!\!\!\!\!\!\!\!\!\!\!\!\! 
S = \int \sqrt{-g}\, d^4x
 \Big\{ 
 \frac{1}{2\kappa^2}\left(R-2\Lambda\right) 
 \nonumber\\
 &&\ \ \ \ \ \ \ \ \ \ \ \ \ \,  +G_{\mu\nu}\left[\alpha 
T^{\mu\nu}+\beta 
(\partial^{\mu}T)(\partial^{\nu}T)\right]+   L_m
 \Big\}  \,,\label{action}
\end{eqnarray}
where $ G_{\mu\nu}$ is the Einstein tensor, $T^{\mu\nu}$ is the energy-momentum 
tensor 
defined as
\begin{align}
T_{\mu\nu}=-\frac{2}{\sqrt{-g}}\frac{\delta(\sqrt{-g}L_{m})}{\delta 
g^{\mu\nu}}\,,
\end{align}
with $L_m$ the matter Lagrangian, and $T=T^{\mu\nu}g_{\mu\nu}$ is its trace. 
Furthermore,   $\kappa^2$ is the gravitational constant, while $\alpha$ and 
$\beta$ are the coupling parameters, while for completeness we consider the 
cosmological constant $\Lambda$ too. Similarly to Horndeski construction
\cite{Horndeski:1974wa,DeFelice:2010nf,Deffayet:2011gz} the use of the Einstein 
tensor ensures that the resulting field equations will not contain higher-order 
derivatives.

Variation of the action with respect to the metric leads to the 
following field equations
\begin{equation} 
G_{\mu\nu}+\Lambda\tensor{g}{_\mu_\nu}=\kappa^2\tilde{T}_{\mu\nu}=\kappa^2[  
T_{\m\n}+\a T_{\m\n}^{(\a)}+\b 
T_{\m\n}^{(\b)}],
 \label{geneqs}
\end{equation}
where we have defined
\begin{eqnarray}
&&
\!\!\!\!\!\!\!\!\!
T_{\m\n}^{(\a)}\equiv\tensor{g}{_\mu_\nu}\tensor{T}{_\alpha_\beta}\tensor{G}{
^\alpha^\beta}+\tensor{R}{_\mu_\nu}T-2\tensor{G}{_\nu^\alpha}\tensor{T}{
_\mu_\alpha}-2\tensor{G}{_\mu^\alpha}\tensor{T}{_\nu_\alpha}
\nonumber  \\
&& 
\ \ \ \ \ -R\tensor{T}{
_\mu_\nu}-\Box\tensor{T}{_\mu_\nu}+\nabla_{\alpha}\nabla_{\mu}\tensor{T}{
_\nu^\alpha}+\nabla_{\alpha}\nabla_{\nu 
} 
\tensor{T}{_\mu^\alpha} \nonumber  \\
&& 
\ \ \ \ \ 
-\tensor{g}{_\mu_\nu}\left(\nabla_{\alpha}\nabla_{\beta} 
\tensor{T}{^\alpha^\beta}\right)+\tensor{g}{_\mu_\nu}\Box 
T-\nabla_{\mu}\nabla_{\nu}T\nonumber  \\
&& 
\ \ \ \ \ -2\tensor{\Xi}{_\mu_\nu}\,, 
\end{eqnarray}
and 
\begin{eqnarray}
&&
\!\!\!
T_{\m\n}^{(\b)}\equiv
\tensor{g}{_\mu_\nu}\tensor{G}{^\alpha^\beta}
\left(\nabla_{\alpha}T\right)\left(\nabla_{\beta}T\right) 
+\tensor{g}{_\mu_\nu}\tensor{R}{^\alpha^\beta}\left(\nabla_{\alpha}
T\right)\left(\nabla_{\beta}T\right) \nonumber  \\
&& 
\ \ \ \ \ \ \ \ 
+\tensor{R}{_\mu_\nu}\left(\nabla_{\alpha}T\right)\left(\nabla^{\alpha}
T\right)-2\left(\nabla_{\alpha}\nabla_{\nu}T\right)\left(\nabla^{\alpha}\nabla_{
\mu}T\right) \nonumber \\
&& \ \ \ \ \ \ \ \ 
+\tensor{g}{_\mu_\nu}\left(\nabla_{\alpha}\nabla_{\beta}T\right)\left(\nabla^{
\alpha}\nabla^{\beta}T\right)
-\tensor{g}{_\mu_\nu}\left(\Box 
T\right)^{2}\nonumber  \\
&& 
\ \ \ \ \ \ \ \  -2\tensor{R}{_\mu_\alpha_\nu_\beta}\left(\nabla^{\alpha}
T\right)\left(\nabla^{\beta}T\right)-2\tensor{G}{_\nu^\alpha}\left(\nabla_{
\alpha}T\right)\left(\nabla_{\mu}T\right) \nonumber \\
&&\ \ \ \ \ \ \ \ 
-2\tensor{G}{_\mu^\alpha}\left(\nabla_{\alpha}T\right)\left(\nabla_{\nu}
T\right)-R\left(\nabla_{\mu}T\right)\left(\nabla_{\nu}T\right)\nonumber  \\
&& 
\ \ \ \ \ \ \ \ 
+2\left(\nabla_{\alpha}\nabla^{\alpha}T\right)\left(\nabla_{\mu}\nabla_{\nu}
T\right)\nonumber  \\
&& 
\ \ \ \ \ \ \ \  +4\tensor{G}{_\alpha_\beta}\nabla^{\alpha}\nabla^{\beta}T\left 
(\tensor{T}{_\mu_\nu}+\tensor{\Theta}{_\mu_\nu}\right ).
\end{eqnarray}
In the above expressions 
 we have used that
\begin{equation}
\frac{\tensor{\delta T}{_\alpha_\beta}}{\tensor{\delta 
g}{^\mu^\nu}}=\frac{\tensor{\delta g}{_\alpha_\beta}}{\tensor{\delta 
g}{^\mu^\nu}}\mathcal{L}_m+\frac{1}{2}\tensor{g}{_\alpha_\beta}\tensor{g}{
_\mu_\nu}\mathcal{L}_m-\frac{1}{2}\tensor{g}{_\alpha_\beta}\tensor{T}{_\mu_\nu}
-2\frac{\partial^2\mathcal{L}_m}{\partial\tensor{g}{^\mu^\nu}\partial\tensor{g}{
^\alpha^\beta}},
\end{equation}
and we have introduced  the tensors $\tensor{\Theta}{_\mu_\nu}$ and 
$\tensor{\Xi}{_\mu_\nu}$   as
\begin{eqnarray}
&& \!\!\!\!\!
\tensor{\Theta}{_\mu_\nu}\equiv  g^{\a\b}\frac{\d T_{\a\b}}{\d 
g^{\m\n}}\nonumber\\
&& \ \ \ \ 
=\tensor{g}{_\mu_\nu}\mathcal{L}_m-2\tensor{T}{_\mu_\nu}-2\tensor{g}{
^\alpha^\beta}\frac{\delta^2\mathcal{L}_m}{\delta\tensor{g}{^\mu^\nu}
\delta\tensor{g}{^\alpha^\beta}}\,, 
\end{eqnarray}
and
\begin{eqnarray}
&& \!\!\!\!\!
\tensor{\Xi}{_\mu_\nu}\equiv  G^{\a\b}\frac{\d T_{\a\b}}{\d 
g^{\m\n}}\nonumber\\
&& \ \ \ \, 
=-\tensor{G}{_\mu_\nu}\mathcal{L}_m+\frac{1}{2}\tensor{G}{^\alpha^\beta
}\tensor{g}{_\alpha_\beta}\left(\tensor{g}{_\mu_\nu}\mathcal{L}_m-\tensor{T}{
_\mu_\nu}\right)
\nonumber\\
&& \ \ \ \ \ \ \, 
-2\tensor{G}{^\alpha^\beta}\frac{\delta^2\mathcal{L}_m}{
\delta\tensor{g}{^\mu^\nu}\delta\tensor{g}{^\alpha^\beta}}.
\label{Xitensor00}
\end{eqnarray}
Note that due to the specific form of action (\ref{action}),
the above general field equations do not 
contain higher-order 
derivatives, and thus the theory does not suffer from ghost instabilities.
Finally, taking the covariant derivative of (\ref{geneqs}), and using that 
$\nabla^\mu G_{\mu\nu}=0$, we can obtain as usual the conservation equation
$\nabla^\mu G_{\mu\nu} = \kappa^2 \nabla^\mu \tilde{T}_{\mu\nu} = 0$, namely
\begin{align}
\nabla^\mu[  
T_{\m\n}+\a T_{\m\n}^{(\a)}+\b 
T_{\m\n}^{(\b)}]= 0. 
\label{genconserveq}
\end{align}
Lastly, we mention that in the case where $\alpha=\beta=0$ we recover General 
Relativity and $\Lambda$CDM cosmology.

 \section{Cosmological Applications} 
 \label{Applications}
 
 In the previous section we constructed theories with couplings between   
the Einstein tensor and the energy-momentum tensor and its trace. In the 
present section we proceed to the investigation of their cosmological 
applications. We consider a flat Friedmann-Robertson-Walker (FRW) spacetime 
metric of the form
\begin{eqnarray}
ds^2=-dt^2+a(t)^2\d_{ij}dx^i dx^j,
\label{metric}
\end{eqnarray}
where $a(t)$ is the scale factor. For the matter sector we consider the 
standard perfect fluid,  with energy-momentum tensor  
\begin{eqnarray}
T_{\m\n}=(\rho_m+p_m)u_\m u_\n+p_m g_{\m\n},
\end{eqnarray}
where $u^\m $ is the 4-velocity which satisfies $u_\m u^\n=-1$. 
Moreover, concerning the matter Lagrangian we assume the standard form  
$\mathcal{L}_{m}=p_m$ 
\cite{Bertolami:2007gv,Bertolami:2008zh,Bertolami:2008ab,Harko:2008qz,
Harko:2010mv,Harko:2012hm,Wang:2012rw,Fadragas:2014mra}.
Under these considerations,   
 the tensors $\tensor{\Theta}{_\mu_\nu}$ and 
$\tensor{\Xi}{_\mu_\nu}$ become
\begin{eqnarray}
\tensor{\Theta}{_\mu_\nu}&=&-2\left(\rho_m+p_m\right)u_{\mu}u_{\nu}-p_m\tensor{g
}{_\mu_\nu}\,,   \\
\tensor{\Xi}{_\mu_\nu}&=&-\tensor{G}{_\mu_\nu}p_m-\frac{1}{2}\tensor{G}{
^\alpha^\beta}\tensor{g}{_\alpha_\beta}\left(\rho_m+p_m\right)u_\mu u_\nu .
\label{Xitensor}
\end{eqnarray}

\subsection{Background Evolution}
\label{Background}

We  start our investigation by the examination of the background evolution.
Substituting (\ref{metric})-(\ref{Xitensor}) into the general field equations 
(\ref{geneqs}) we obtain the Friedmann equations 
\begin{eqnarray}
3H^2-\Lambda&=&\k^2 (  \r_m+\r_\a+\r_\b)\,,
\label{Fr1}\\
3H^2+2\dot{H}-\Lambda&=&-\k^2(  p_m+p_\a+p_\b)\,,
\label{Fr2}
\end{eqnarray} 
with $H=\dot{a}/a$ the Hubble function, and where we have introduced
\begin{eqnarray}
&&
\!\!\!\!\!\!\!
\r_\a\equiv\alpha \left [-6(\pm+\rm)\hd+3H^2(\pm-\rm)+6H\dot{p}_m\right ]\,, 
\nonumber \\
&&\!\!\!\!\!\!\!\r_\b
\equiv\beta\Big\{-12H(\pm+\rm)(3\dot{p}_{m}-\dot{\r}_m)(3H^2+2\hd)
\nonumber\\
&& \ \ \ \ \ \ \ \  
-3
H^2\left[4(\pm+\rm)(3\ddot{p}_{m}-\ddot{\r}_{m})\right.\nonumber\\
&& \ \ \ \ \   \ \ \  \ \ \ \ \ \ \ \ \  \left. -3(3\dot{p}_{m}-\dot{\r}_{m}
)^2\right ]\!\Big \}\,, 
\end{eqnarray}
and 
\begin{eqnarray}
&&
\!\!\!\!\!\!\!
p_\a\equiv\alpha\Big
\{-(3\pm+\rm)(3H^2+2\hd)
\nonumber\\
&& \ \ \ \ \ \ \ \ \,
-\left[2H(3\dot{p}_{m}+\dot{\r}_{m})+2\ddot{p}_{m}\right
] \Big\}\, , \nonumber \\
&&
\!\!\!\!\!\!\!
p_\b\equiv\beta\Big \{-(3\dot{p}_{m}-\dot{\r}_{m})^2(3H^2+2\hd)
\nonumber\\
&& \ \ \ \ \ \ \ \ \,
-4 
H(3\dot{p}_{m}-\dot{\r}_{m})(3\ddot{p}_{m}-\ddot{\r}_{m})\Big\}\,.
\end{eqnarray}
Additionally, in the case of FRW metric,  the general conservation 
\eqref{genconserveq}  becomes
\begin{equation}
 \dot{\rho}_m +\dot{\rho}_\alpha 
+\dot{\rho}_\beta+3H\left( \rho_m+\rho_\alpha+\rho_\beta+  
p_m+p_\alpha+p_\beta
\right)=0.
\label{coserveq11}
\end{equation}

 We mention here that, as expected by the form of the general field 
equations (\ref{geneqs})-(\ref{Xitensor00}), the Friedmann equations 
(\ref{Fr1}),(\ref{Fr2}) do not contain higher-order 
derivatives, and thus the theory is free from
Ostrogradsky instabilities. Nevertheless, in the theory at hand, 
although the field equations are healthy, one can 
in principle have  higher than second order derivatives in the general 
conservation equation (\ref{coserveq11}) (in particular in the $\beta$-term). 
This is typical in all theories  of 
non-minimal matter couplings (see e.g. the relevant discussion in the 
well-known paper \cite{Harko:2011kv}). However, this is not a problem, since the 
conservation equation is not used in order to obtain the solutions (one 
uses only the two Friedmann equations),  and after one has extracted the 
solutions he inserts them in the conservation equation which is trivially 
satisfied as expected.

As we observe, in the scenario of non-minimal coupling between the matter 
sector and the Einstein tensor, we obtain an effective dark energy sector with 
energy density and pressure 
\begin{eqnarray}
 &&\rho_{DE}\equiv\frac{\Lambda}{\kappa^2}+\rho_\alpha+\rho_\beta\\
 &&p_{DE}\equiv-\frac{\Lambda}{\kappa^2}+p_\alpha+p_\beta,
\end{eqnarray}
respectively, and thus with effective equation-of-state parameter 
\begin{eqnarray}
w_{DE}\equiv \frac{p_{DE}}{\rho_{DE}}.
\end{eqnarray}
 Note that the above total conservation equation (\ref{coserveq11}) can be 
further handled in two ways. 
The first is to consider that although the total energy is conserved 
the individual sectors do not,  
namely we obtain an effective interaction and a transfer of energy between   
matter and geometric sectors and vice versa, typical in all theories with 
matter-geometry couplings 
\cite{Bertolami:2007gv,Bertolami:2008zh,Bertolami:2008ab,Harko:2008qz,
Harko:2010mv,Harko:2012hm,Wang:2012rw,Fadragas:2014mra,Harko:2011kv,
Momeni:2011am,Sharif:2012zzd,Alvarenga:2013syu,Shabani:2013djy,Noureen:2015xva,
Zaregonbadi:2016xna,Haghani:2013oma,Odintsov:2013iba,Harko:2014sja,
Carloni:2015lsa,Gonzalez-Espinoza:2018gyl,
Harko:2014aja,Junior:2015bva,Saez-Gomez:2016wxb, 
Farrugia:2016pjh,Pace:2017dpu} (this effective interaction, apart from other 
features, has the advantage that 
it can alleviate the coincidence problem). The second choice is to 
additionally impose by hand the standard matter conservation equation, and thus 
to obtain also a separate  conservation equation for the effective dark energy 
sector. In the following without loss of generality we make the first 
choice, and hence separately the matter sector is not 
conserved, however due to the imposed parameter values the effective 
interaction between dark-matter and dark energy is weak, and thus matter scales 
very close to $a(t)^{-3}$ as required by observations.

Let us proceed to the numerical examination of the above Friedmann equations.
Without loss of generality we  focus on the case of dust matter 
($p_m\approx0$). Moreover, we introduce the  density 
parameters as 
 \begin{eqnarray} \label{FRWomatter}
&&\Omega_m\equiv\frac{\kappa^2}{3H^2} \rho_m\\
&& \label{FRWode}
\Omega_{DE}\equiv\frac{\kappa^2}{3H^2} \rho_{DE},
 \end{eqnarray} 
for the matter and effective dark energy sector respectively. Finally, as usual 
we use the redshift  $ 
 1+z=a_0/a$ as the independent variable, setting the present scale factor to 
$a_0=1$ (from now on the subscript ``0'' denotes the   value of a quantity at 
present).  

\begin{figure}[!h]
\centering
\includegraphics[width=9cm]{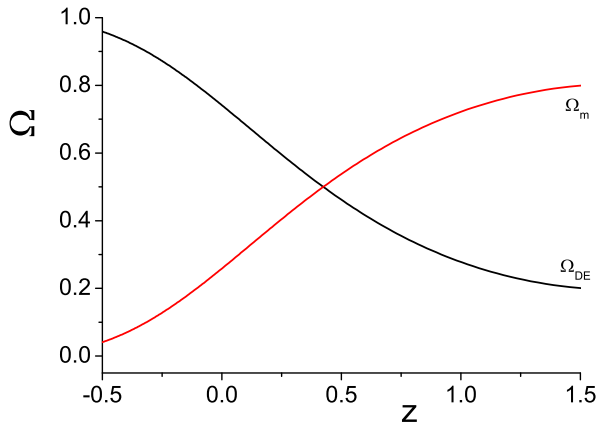}                                   
\caption{\it{ The evolution of the effective dark energy density 
parameter $\Omega_{DE}$ (black-solid) and of the matter density parameter 
$\Omega_{m}$ (red-dashed),   as a function of the redshift $z$, for the 
scenario of non-minimal  (derivative) coupling between matter and Einstein 
tensor, with $\alpha=-0.05$ and $\beta=0.01$, in units where $\kappa^2=1$.  We 
have imposed $ \Omega_{m0}\approx0.3$ at present time. 
}}
\label{Omegas}
\end{figure}
\begin{figure}[!h]
\centering
\includegraphics[width=9cm]{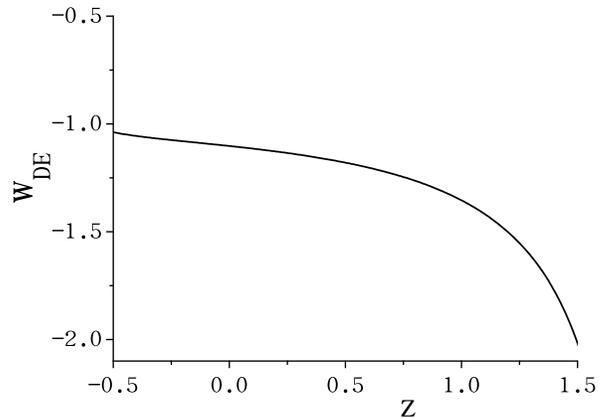}
\caption{\it{The evolution of the effective dark-energy equation-of-state 
parameter $w_{DE}$, as a function of the redshift $z$, for the 
scenario of non-minimal  (derivative) coupling between matter and Einstein 
tensor, with $\alpha=-0.05$ and $\beta=0.01$, in units where $\kappa^2=1$.  We 
have imposed $ \Omega_{m0}\approx0.3$ at present time. }}
\label{figwDE}
\end{figure}

We solve equations   (\ref{Fr1}),(\ref{Fr2}) numerically, imposing  the  
conditions
$\Omega_{DE}(z=0)\equiv\Omega_{DE0}\approx0.7$ and therefore 
$\Omega_m(z=0)\equiv\Omega_{m0}\approx0.3$ in agreement with
observations \cite{Planck:2018vyg}, which then determines the relation between 
$\alpha$, $\beta$ and $\Lambda$. In Fig. \ref{Omegas} we draw the resulting 
evolution of $\Omega_{DE}(z)$ and $\Omega_{m}(z)$.
As we can see the scenario at hand can  describe the thermal  history of the 
universe successfully, namely the sequence of matter and late-time acceleration 
epochs.  
Additionally, in 
Fig. \ref{figwDE} we depict the corresponding evolution of the effective 
equation-of-state parameter.  As we can see,  $w_{DE}$ is algebraically smaller 
in the past, while it is closer to  $-1$ at present time, as required 
  by observations, before going asymptotically to $-1$ in the future where the 
cosmological constant dominates. Note that in this example  $w_{DE}$ lies 
in the phantom regime despite the fact that the effective dark-energy sector 
constitutes from dust matter terms. This is  typical also in other 
scenarios of couplings between matter and geometry mentioned in the 
Introduction, and reveals the capabilities of such interactions  
\cite{Bertolami:2007gv,Bertolami:2008zh,Bertolami:2008ab,Harko:2008qz,
Harko:2010mv,Harko:2012hm,Wang:2012rw,Fadragas:2014mra,Harko:2011kv,
Momeni:2011am,Sharif:2012zzd,Alvarenga:2013syu,Shabani:2013djy,Noureen:2015xva,
Zaregonbadi:2016xna,Harko:2014sja,Carloni:2015lsa,Gonzalez-Espinoza:2018gyl,
Harko:2014aja,Junior:2015bva,Saez-Gomez:2016wxb, Farrugia:2016pjh,Pace:2017dpu}.

\begin{figure}[ht]
	\centering
	\includegraphics[width=0.5\textwidth]{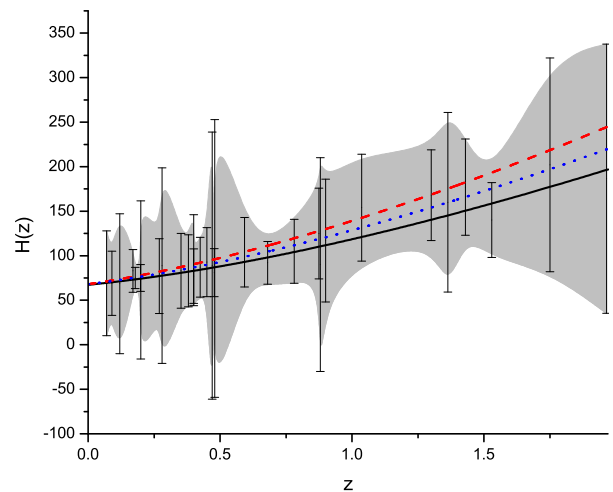}
	\caption{{\it{ The $H(z)$ in units of Km/s/Mpc as a 
 function of the redshift,  for or the 
scenario of non-minimal  (derivative) coupling between matter and Einstein 
tensor, with $\alpha=-0.05$ and $\beta=0.01$ (red-dashed), and with 
$\alpha=-0.01$ and $\beta=0.001$ (blue-dotted) in units where $\kappa^2=1$, on  
 top of the Cosmic Chronometers data points from \cite{Yu:2017iju} at $3\sigma$ 
confidence level. For comparison  we also
present the $\Lambda$CDM  curve  (black - solid). We have imposed 
				$\Omega_{m0}\approx0.3$. }} }
	\label{datafig}
\end{figure}

For consistency purposes, we close this subsection by a brief confrontation 
with  the Cosmic Chronometer 
  datasets, which are based on the $H(z)$  measurements  through the 
relative ages of  massive and passively evolving galaxies and the 
corresponding estimation of  $dz/dt$   \cite{Jimenez:2001gg}. In Fig.   
\ref{datafig}   we   compare the $H(z)$ evolution predicted from our scenario, 
 with the $H(z)$ Cosmic Chronometer  data from \cite{Yu:2017iju} at 
$3\sigma$ confidence level, while for completeness we present the $\Lambda$CDM 
curve too. As we observe  the agreement is very good, and the $H(z)$ 
evolution  lies within the errors of the direct 
measurements  of $H(z)$, exhibiting a slightly higher accelerating behavior in 
the past due to the phantom nature of dark energy in these examples.

\subsection{Cosmological perturbations}
 \label{perturbations}

 In this subsection we perform a detailed perturbation analysis of the theory 
at 
hand in the linear regime 
\cite{Mukhanov:1990me,Ma:1995ey, 
Linder:2005in,Uzan:2006mf,Basilakos:2013nfa,Nesseris:2013jea,
Basilakos:2014yda}. 
 Concerning the scalar perturbations we start form 
the standard perturbed 
metric of isentropic   perturbations   in the Newtonian gauge: 
\be
ds^2=-\left(1+2\phi\right)dt^2+a^2\tensor{\delta}{_i_j}\left(1-2\psi\right)dx^{i
}dx^{j},
\ee
and as usual we consider the expressions
\begin{eqnarray}
&&\delta\tensor{T}{^0_0}=-\delta\rho_{m}
\nonumber\\
&&\delta\tensor{T}{^i_j}=\delta p_{m}\tensor{\delta}{_i_j}
\nonumber\\
&&\delta\tensor{T}{^i_0}=-\frac{\left(\rho_{m}+p_{m}\right)}{a}\partial_iV
\nonumber\\
&&\delta\tensor{T}{^0_i}=a\left(\rho_{m}+p_{m}\right)\partial_iV,
\end{eqnarray}
where $V$ is defined through $u_{\mu}=a\left(-\phi,\partial_iV\right) $. 
Hence, inserting these in the general field equations (\ref{geneqs}), and using 
additionally the background Friedmann equations  (\ref{Fr1}),(\ref{Fr2}) to 
eliminate terms, we finally obtain the time-time and the   space-diagonal 
equations respectively given by 
\begin{widetext}
 \begin{eqnarray}
&&
6H(H\phi+\dot{\psi})-2\frac{\nabla^2}{a^2}\psi
=
-\kappa^2\Big\{ 
\delta\rho_{m}+\alpha\Big\{3(H^2-2\dot{H})\delta 
p_{m}+6H\delta\dot{p}_{m}-3(H^2+2\dot{H})\delta\rho_{m}+6\left(\rho_{m}+p_{m}
\right)\ddot{\psi}
\nonumber 
\\
&&\left.\left.
+6\left[H^2\left(\rho_{m}-p_{m}\right) 
 -2H\dot{p}_{m}+2\dot{H}\left(\rho_{m}+p_{m}\right)\right]
\phi+6H\left(\rho_{m}+p_{m}\right)\dot{\phi}+6\left[H\left(\rho_{m}-p_{m}
\right)-\dot{p}_{m}\right]\dot{\psi}
\right.\right. \nonumber \\
&&\left.\left. 
+\frac{\nabla^2}{a^2}\left[4H\left(\rho_{m}+p_{m}\right)aV+2\left(\rho_{m}+p_{m}
\right)\phi-2\left(\rho_{m}-p_{m}\right)\psi- 2\delta 
p_{m}\right]\Big\}\right.\right. 
\nonumber \\
&&\left. 
+\beta
\Big\{-36H^2\left(\rho_{m}+p_{m}\right)\delta\ddot{p}_{m}
-18[
6H^3\left(\rho_{m}+p_{m}\right)+4H\dot{H}\left(\rho_{m}+p_{m}
\right)+3H^2\left(\dot{\rho}_{m}-3\dot{p}_{m}\right)]\delta\dot{p}_{m}
\right. 
\nonumber \\
&&
+12H \left[
 (2\dot{H}+3H^2 )\left(\dot{\rho}_{m}-3\dot{p}_{m}
\right)+H\left(\ddot{\rho}_{m}-3\ddot{p}_{m}\right)\right ]
\left(\delta\rho_{m}+\delta 
p_{m}\right)\nonumber\\
&&\left.
\left.
+6H
[(4\dot{H}
+6H^2)\left(\rho_{m}+p_{m}\right)+3H\left(\dot{\rho}_{m}-3\dot{p}_{m}
\right)]
\delta\dot{\rho}_{m}+12H^2\left(\rho_{m}+p_{m}\right)\delta\ddot{\rho}_{m}
\right.\right. \nonumber \\
&&\left.\left.+
\left\{
 (144H^3+96H\dot{H} )\left[3p_{m}\dot{p}_{m}+\rho_{m}\left(3\dot{p}_{m}
-\dot{\rho}_{m}\right)-p_{m}\dot{\rho}_{m}\right]\right.\right.\right. \nonumber 
\\ 
&&\left.\left.\left.\left.
+12H^2\left[12p_{m}\ddot{p}_{m}-3\left(3\dot{p}_{m}
-\dot{\rho}_{m}\right)^{2}+4\rho_{m}\left(3\ddot{p}_{m}-\ddot{\rho}_{m}
\right)-4p_{m}\ddot{\rho}_{m}\right]\right\}
\phi\right.\right.\right. \nonumber \\
&&\left.+36 
H^2\left[3p_{m}\dot{p}_{m}+\rho_{m}\left(3\dot{p}_{m}-\dot{\rho}_{m}\right)-p_{m
}\dot{\rho}_{m}\right]\dot{\phi}
+\left\{4(27H^2+6\dot{H})\left[3p_{m}\dot{p}_{m}+\rho_{m}
\left(3\dot{p}_{m}-\dot{\rho}_{m}\right)-p_{m}\dot{\rho}_{m}\right]
\right.\right. \nonumber \\
&&\left.\left.\left.+6H\left[12p_{m}\ddot{p}_{m}-3\left(3\dot{p}_{m}-\dot{\rho}_
{m}\right)^{2}+4\rho_{m}\left(3\ddot{p}_{m}-\ddot{\rho}_{m}\right)-4p_{m}\ddot{
\rho}_{m}\right]\right\}
\dot{\psi}+3H\left[24p_{m}\dot{p}_{m}+8\rho_{m}
\left(3\dot{p}_{m}-\dot{\rho}_{m}\right)-8p_{m}\dot{\rho}_{m}\right]\ddot{\psi}
\right.\right. \nonumber \\
&&\left.\left.
+4\left[(2\dot{H}+3H^2)\left(\rho_{m}+p_{m}
\right)+H\left(\dot{\rho}_{m}-3\dot{p}_{m}\right)\right]\frac{\nabla^2}{a^2}
\left(3\delta 
p_{m}-\delta\rho_{m}\right)+8H\left(\rho_{m}+p_{m}\right)\left(3\dot{p}_{m}-\dot
{\rho}_{m}\right)\frac{\nabla^2}{a^2}\phi\right.\right. \nonumber \\ 
&&
 +2\left[\left(3\dot{p}_{m}-\dot{\rho}_{m}\right)^{2}-12p_{m}\ddot{
p}_{m}+4H\left(\rho_{m}+p_{m}\right)\left(\dot{\rho}_{m}-3\dot{p}_{m}\right)+4p_
{m}\ddot{\rho}_{m}+4\rho_{m}\left(\ddot{\rho}_{m}-3\ddot{p}_{m}\right)\right]
\frac{\nabla^2}{a^2}\psi\Big\}
\Big\},
\label{eq1}
\end{eqnarray} 
\begin{eqnarray}
&& 
2\left(3H^2+2\dot{H}\right)\phi+2H\left(\dot{\phi}+3\dot{\psi}\right)+2\ddot{
\psi}+\frac{\nabla^2}{a^2}\left(\phi-\psi\right)=\kappa^2
\Big\{ \delta 
p_{m} \nonumber \\
&&\left. 
-\alpha\Big\{2\delta\ddot{p}_{m}-4\ddot{p}_{m}\phi 
+\left(3H^2+2\dot{H}\right)\left[3\delta 
p_{m}+\delta\rho_{m}-2\left(\rho_{m}+3p_{m}\right)\phi\right]-2\dot{\rho}\dot{
\psi}-2\ddot{\psi}\left(\rho_{m}+3p_{m}\right) \right.   \nonumber \\
&&\left.\left.+2H\left[3\delta\dot{p}_{m}+\delta\dot{\rho}_{m}-2\left(\dot{\rho}
_{m}+3\dot{p}_{m}\right)\phi-\left(\rho_{m}+3p_{m}\right)\left(\dot{\phi}+3\dot{
\psi}\right)\right]-2\dot{p}_{m}\left(\dot{\phi}+3\dot{\psi}
\right)\right.\right. \nonumber \\
&&\left.\left. 
+\frac{\nabla^2}{a^2}
\Big\{2\left[\dot{\rho}_{m}+\dot{p}_{m}+2H\left(\rho_{m}+p_
{m}\right)\right]aV+2\left(\rho_{m}+p_{m}\right)a\dot{V}-\delta 
p_{m}+\delta\rho_{m}+\left(\rho_{m}-p_{m}\right)\left(\phi-\psi\right)\Big\}
\Big\}
\right.\right. \nonumber \\
&&\left.\left.-\beta
\Big\{2\left(2\dot{H}+3H^2\right)\left[3\dot{p}_{m}
\left(3\delta\dot{p}_{m}-\delta\dot{\rho}_{m}\right)-2\phi\left(9\dot{p}_{m}^{2}
+\dot{\rho}_{m}^{2}\right)+\dot{\rho}_{m}\left(\delta\dot{\rho}_{m}-3\delta\dot{
p}_{m}+12\dot{p}_{m}\phi\right)\right]\right.\right.  \nonumber \\
&&\left.\left.-12\dot{p}_{m}\left(3\ddot{p}_{m}-\ddot{\rho}_{m}\right)\dot{\psi}
+4H
\Big\{3\ddot{p}_{m}\left(3\delta\dot{p}_{m}-\delta\dot{\rho}_{m}\right)+\ddot
{\rho}_{m}\left(\delta\dot{\rho}_{m}-3\delta\dot{p}_{m}+12\dot{p}_{m}
\phi\right)+\dot{p}_{m}\left(9\delta\ddot{p}_{m}-3\delta\ddot{\rho}_{m}-36\ddot{
p}_{m}\phi\right)\right.\right.  \nonumber \\
&& \left.\left.-\frac{3}{2}\left(9\dot{p}_{m}^{2}+\dot{\rho}_{m}^{2}
\right)\left(\dot{\phi}+\dot{\psi}\right)+\dot{\rho}_{m}\left[\delta\ddot{\rho}_
{m}-3\delta\ddot{p}_{m}+4\left(3\ddot{p}_{m}-\ddot{\rho}_{m}\right)\phi+9\dot{p}
_{m}\left(\dot{\phi}+\dot{\psi}\right)\right]\Big\}
\right.\right. \nonumber \\
&& -2\left(9\dot{p}_{m}^2+\dot{
\rho}_{m}^{2}\right)\ddot{\psi} 
+4\dot{\rho}_{m}\left[\left(3\ddot{p}_{m}-\ddot{\rho}_{m}
\right)\dot{\psi}+3\dot{p}_{m}\ddot{\psi}\right]
\nonumber \\
&& 
+2\frac{\nabla^2}{a^2}
\Big\{
\left[H\left(\dot{\rho}_{m}-3\dot{p}_{m}\right)+\ddot{\rho}_{m}-3\ddot{p}_{m}
\right]\left(3\delta 
p_{m}-\delta\rho_{m}\right)-\left(3\dot{p}_{m}-\dot{\rho}_{m}\right)^{2}
\left(\phi+\psi\right)
\Big\}\Big\}\Big\},
\label{eq2}
\end{eqnarray}
 \end{widetext}
while  the space non-diagonal equation reads as
\begin{eqnarray}
&&
\!\!\!\!\!\!\!\!
\psi-\phi=-\kappa^2\Big\{\alpha\Big\{\delta 
p_{m}-\delta\rho_{m}\nonumber\\
&&
-2a\left[2H\left(\rho_{m}+p_{m}\right)+\dot{p}_{m
} +\dot { \rho
}_{m}\right]V
\nonumber\\
&&
-2a\left(\rho_{m}+p_{m}\right)\dot{V}+\left(\rho_{m}-p_{m}
\right)\left(\psi-\phi\right)\Big\} \nonumber \\
&& +\beta\Big\{2\left[H\left(3\dot{p}_{m}-\dot{\rho}_{m}\right)+3\ddot{p}_{
m}-\ddot{\rho}_{m}\right]\left(3\delta 
p_{m}-\delta\rho_{m}\right)\nonumber\\
&&
+\left(3\dot{p}_{m}-{\rho}_{m}\right)^{2}
\left(\phi+\psi\right)\Big\}\Big\}.
\label{eq3}
\end{eqnarray}
Note that the last equation implies that the anisotropic stress in the scenario 
at hand is non-zero, and it becomes zero in the case $\alpha=\beta=0$.

For completeness, we investigate the tensor perturbations, too. As usual we 
consider
\be
ds^2=-dt^2+a^2\left(\tensor{\delta}{_i_j}+2\tensor{h}{_i_j}\right)dx^{i}dx^{j},
\ee
 where $\tensor{h}{_i_j}$ is transverse and traceless. Hence, we finally obtain 
the tensor equation 
\begin{eqnarray}
&&\!\!\!\!\!\!\! \!\!\!\!\!\!\!\!\!\!\! 
\tensor{\ddot{h}}{_i_j}+3H\tensor{\dot{h}}{_i_j}-\frac{\nabla^2}{a^2}\tensor{h
}{_i_j}=\kappa^2
\Big\{\alpha
\Big\{
 \left(\rho_{m}+3p_{m}
\right)\tensor{\ddot{h}}{_i_j}\nonumber\\
&&\ \, \ \ 
+
\left[3H\left(\rho_{m}+3p_{m}\right)+\dot{
\rho}_{m}+3\dot{p}_{m}\right]\tensor{\dot{h}}{_i_j} 
\nonumber\\
&&\ \, \ \ 
+\left(\rho_{m}\!-\!p_{m}\right)\frac{\nabla^2}{a^2}
\tensor{h}{_i_j} \!\Big\} \nonumber \\
&&\ \, \ \ 
+\beta\Big\{
 \left(3\dot{p}_{m}-{\rho}_{m}
\right)^{2}\tensor{\ddot{h}}{_i_j}
 +\left(3\dot{p}_{m}-{\rho}_{m}\right)^{2}\frac
{\nabla^2}{a^2}\tensor{h}{_i_j}
\nonumber \\
&&\ \ \ \ \ \ \ \ \ 
+
\left[3H\left(3\dot{p}_{m}-{\rho}_{m}\right)^{2}+6\dot{p}_{m
}\left(3\ddot{p}_{m}-\ddot{\rho}_{m}\right)\right.
\nonumber \\
&&\ \ \ \ \ \ \  \  \ \ \ \ \ \left. +2\dot{\rho}_{m}\left(\ddot{\rho}_{m}
-3\ddot{p}_{m}\right)\right]\tensor{\dot{h}}{_i_j}
 \Big\}\Big\}.
\end{eqnarray}
 
Let us now focus on the scalar perturbations, and introduce as usual the 
important quantity   $\delta\equiv \frac{\delta\rho_m}{\rho_m}$, namely the 
matter overdensity. Inserting $\delta$ into (\ref{eq1})-(\ref{eq3}),
considering dust matter, and transforming as usual to the Fourier space, with 
$k$  the wavenumber, we   extract the evolution equation for $\delta$, which 
can be solved numerically. However, in order to further simplify the 
expressions, we assume small deviations from $\Lambda$CDM 
cosmology, namely $\kappa^2(\rho_\alpha+\rho_\beta)\ll\Lambda$, which implies 
small values for $\alpha$ and $\beta$ ($\alpha,\beta\ll (\kappa^2H^2)^{-1}$), 
and moreover we focus   on sub-horizon scales, i.e. with $k\gg aH$,
and for the sound speed we consider 
 $ c_{\mathrm{eff}}^{(m)2}\equiv  \delta  p_m/\delta\rho_m\ll1 $. Hence, 
we result in the following equation:
\begin{widetext}
\be 
\ddot{\delta}+2H\dot{\delta}+\frac{\kappa^2  \rho_{m}\delta\left \{6\beta 
\rho_{m}\dot{\rho}_{m}H^{2}+\left(k^{2}+8\beta 
\rho_{m}\dot{H}\right)\dot{\rho}_{m}+H\left[4\beta 
\dot{\rho}_{m}^{2}+3\rho_{m}\left(\alpha -2\beta 
\ddot{\rho}_{m}\right)\right]\right \}}{2k^{2}\dot{\rho}_{m}\left[-1+\kappa^2 
\beta\left(3H\rho_{m}\dot{\rho}_{m}+\dot{\rho}_{m}^{2}+4\rho_{m}\ddot{\rho}_{m} 
\right)\right]}=0.
\label{eqdelta1}
\ee
\end{widetext}
The above equation determines the evolution of matter overdensity in the 
scenario at hand. We can bring it in the more convenient form \cite{
Linder:2005in,Uzan:2006mf,Basilakos:2013nfa,Nesseris:2013jea,
Basilakos:2014yda}
\be 
\ddot{\delta}+2H\dot{\delta}-4\pi G_{\mathrm{eff}}\rho_{m}\delta=0,
\ee
by defining an effective gravitational constant  
\begin{widetext}
\be 
8\pi G_{\mathrm{eff}}=\kappa^2_{\mathrm{eff}}\equiv \frac{\kappa^2  
 \left \{6\beta 
\rho_{m}\dot{\rho}_{m}H^{2}+\left(k^{2}+8\beta 
\rho_{m}\dot{H}\right)\dot{\rho}_{m}+H\left[4\beta 
\dot{\rho}_{m}^{2}+3\rho_{m}\left(\alpha -2\beta 
\ddot{\rho}_{m}\right)\right]\right \}}{ k^{2}\dot{\rho}_{m}\left[1-\kappa^2 
\beta\left(3H\rho_{m}\dot{\rho}_{m}+\dot{\rho}_{m}^{2}+4\rho_{m}\ddot{\rho}_{m} 
\right)\right]}.
\ee
\end{widetext}
As expected, when the nonminimal derivative coupling switches off, i.e. in the 
case $\alpha=\beta=0$, we obtain $\kappa^2_{\mathrm{eff}}=\kappa^2$ and thus 
$G_{\mathrm{eff}}=G$ and hence we recover the results of $\Lambda$CDM cosmology.

\begin{figure}[ht]
	\centering
	\includegraphics[width=0.44\textwidth]{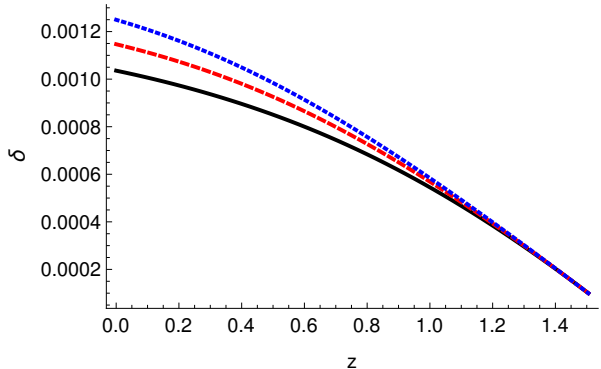}
	\caption{{\it{ The evolution of the matter overdensity   $\delta\equiv 
\frac{\delta\rho_m}{\rho_m}$, as a function of the redshift $z$, for the 
scenario of non-minimal  (derivative) coupling between matter and Einstein 
tensor, with $\alpha=-0.1$, $\beta=0.1$ (black-solid), $\alpha=-0.2$, 
$\beta=0.2$ (red-dashed), and $\alpha=-0.5$, $\beta=0.4$ (blue-dotted),
in units where $\kappa^2=1$, at  scale $k = 10^{-3} $hMpc$^{-1}$.  We 
have imposed $ \Omega_{m0}\approx0.3$ at present time. }} }
	\label{deltaplot}
\end{figure}

In order to provide a more transparent picture of the behavior of 
perturbations and the features of the large scale structure, in Fig. 
\ref{deltaplot} we 
depict the evolution of $\delta$ as a function of the redshift. As we observe, 
the theories with nonminimal (derivative) coupling 
between  matter and the Einstein tensor can describe the evolution of the large 
scale structure in agreement with observations. Furthermore, this evolution is 
sensitive to the coupling parameters, hence one can use $f\sigma_8$ data in 
order to extract constraints on them and break possible degeneracies that may 
appear at the background level.

\section{Conclusions}
\label{Conclusions}

In this work, inspired by the coupling of scalar fields to the 
Einstein tensor, as well as by the coupling of the matter sector to the Ricci 
scalar, we  constructed new classes of modified theories in 
which the matter sector couples with the Einstein tensor. In particular, we 
considered a direct coupling of the energy-momentum tensor to the Einstein 
tensor, and a coupling of the Einstein tensor to the derivatives of the   
trace of the   energy-momentum tensor.

Firstly, we extracted the general field equations, which comparing to General 
Relativity include corrections depending on the two coupling parameters of the 
theory. Then we proceeded to the cosmological application around a flat FRW 
background, and at the background level we extracted the Friedmann equations 
from the general field equations, whose extra terms  can be absorbed in an 
effective dark energy sector.

Assuming the matter sector to be dust we elaborated the equations numerically, 
and we saw that   the scenario at hand can successfully describe  the usual 
thermal history of the universe, with  the sequence of matter and dark-energy 
epochs. Additionally, we examined  the dark-energy equation-of-state 
parameter and we showed that it can lie in the phantom regime, tending 
progressively to $-1$ at present and future times. It is interesting    to 
mention
that this behavior is obtained although the effective dark-energy sector 
constitutes from matter terms, nevertheless it is not uncommon in theories 
with couplings between matter and geometry. Finally, for completeness we 
confronted the theory with     Cosmic Chronometer 
  data, showing that  the agreement is very good, and that the predicted $H(z)$ 
evolution  lies within the errors of the direct 
measurements  of $H(z)$, exhibiting a slightly higher accelerating behavior in 
the past due to the phantom nature of dark energy.

We proceeded to the detailed investigation of the perturbations, both scalar 
and tensor ones. Focusing on   scalar perturbations, we extracted the 
evolution equation of the matter overdensity, which is a crucial observable 
since it quantifies the matter clustering and the large scale structure. 
Elaborating the equation numerically, we saw that the predicted evolution of 
the matter overdensity is in agreement with observations.

It would be both interesting and necessary to  perform a full observational 
confrontation with joined datasets from  Cosmic Microwave Background (CMB), 
Baryonic Acoustic Oscillations (BAO), Supernovae Type Ia (SNIa), and Redshift 
space distrotion (RSD) f$\sigma_8$ observations, in order to extract constraints 
on the new coupling parameters. Additionally, one could perform a detailed 
phase-space analysis in order to investigate the asymptotic behavior of the 
scenario, independently of the initial conditions and the specific evolution of 
the Universe. Moreover, it would be interesting to examine the possible 
alleviation of the $H_0$ and $S_8$ tensions \cite{Abdalla:2022yfr} in these 
theories (the fact that we obtain effective phantom behavior is a promising 
feature).  Finally,  apart from the linear perturbation analysis, whose 
equations   have been presented in this work, we should investigate the the 
non-linear perturbations too, since they play a crucial role when 
gravitational instability grows enough to understand the formation of Large 
Scale Structure. 
Such studies lie beyond the scope of the present work and are 
left for future projects.

\begin{acknowledgments}
This research is co-financed by Greece and the European Union (European Social 
Fund-ESF) through the Operational Programme ``Human Resources Development, 
Education and Lifelong Learning'' in the context of the project
``Strengthening Human Resources Research Potential via Doctorate Research'' 
(MIS-5000432), implemented by the
State Scholarships Foundation (IKY). Additionally, it is partially 
financed by the Basic Research program
of the National Technical University of Athens (NTUA,
PEVE) 65232600-ACT-MTG: ``Alleviating Cosmological
Tensions Through Modified Theories of Gravity''.
The authors would like to acknowledge the contribution of the COST Action 
CA21136 ``Addressing observational tensions in cosmology with systematics and 
fundamental physics (CosmoVerse)''. 

\end{acknowledgments}

\end{document}